\documentclass[useAMS,usenatbib]{mn2e}
\usepackage{graphicx,amsmath,multirow,amssymb}
\usepackage{subfig}
\usepackage{natbib}
\newcommand{\comment}[1]{}

\def\simgt{\lower.5ex\hbox{$\; \buildrel > \over \sim \;$}}
\def\simlt{\lower.5ex\hbox{$\; \buildrel < \over \sim \;$}}

\title[PNe in the LMC]{A test for asymptotic giant branch evolution theories:
Planetary Nebulae in the Large Magellanic Cloud}
\author[Ventura et al.]{P. Ventura$^1$, L. Stanghellini$^{2}$, F. Dell'Agli$^{1,3}$,  
D. A. Garc\'{\i}a--Hern\'andez$^{4,5}$, 
\newauthor
M. Di Criscienzo$^1$  \\
$^1$INAF -- Osservatorio Astronomico di Roma, Via Frascati 33, 00040, Monte Porzio Catone (RM), Italy \\
$^2$National Optical Astronomy Observatory, 950 N. Cherry Avenue, Tucson (AZ) 85719, USA \\
$^3$Dipartimento di Fisica, Universit\`a di Roma ``La Sapienza'', P.le Aldo Moro 5, 00143, Roma, Italy \\
$^{4}$Instituto de Astrof\'{\i}sica de Canarias, E-38205 La Laguna, Tenerife, Spain \\
$^{5}$Departamento de Astrof\'{\i}sica, Universidad de La Laguna (ULL), E-38206 La Laguna, Tenerife, Spain\\
}

\begin{document}

\date{Accepted, Received; in original form }

\pagerange{\pageref{firstpage}--\pageref{lastpage}} \pubyear{2012}

\maketitle

\label{firstpage}

\begin{abstract}
We used a new generation of asymptotic giant branch (AGB) stellar models that include 
dust formation in the stellar winds to find the links between evolutionary models 
and the observed properties of a homogeneous sample of Large Magellanic Cloud (LMC) 
planetary nebulae (PNe). 
Comparison between the evolutionary yields of elements such as CNO and the corresponding 
observed chemical abundances is a powerful tool to shed light on evolutionary 
processes such as hot bottom burning (HBB) and third dredge-up (TDU). 
We found that the occurrence of HBB is needed to interpret the 
nitrogen-enriched ($\log({\rm N/H})+12>8$) PNe.
In particular, N-rich PNe with the lowest carbon content 
are nicely reproduced by AGB models of mass $M \geq 6~M_{\odot}$, whose surface chemistry 
reflects the pure effects of HBB.
PNe  with $\log({\rm N/H})+12<7.5$ correspond to ejecta of stars that have not 
experienced HBB, with initial mass below $\sim$$3~M_{\odot}$. Some of these stars show
very large carbon abundances, owing to the many TDU episodes experienced.
We found from our LMC PN sample that there is a threshold to the amount of carbon 
accumulated at AGB surfaces, $\log({\rm C/H})+12<9$. 
Confirmation of this constraint would indicate that, after the C-star stage is reached,
AGBs experience only a few thermal pulses, which suggests a 
rapid loss of the external mantle, probably owing to the effects of radiation pressure on
carbonaceous dust particles present in the circumstellar envelope.
The implications of these findings for AGB evolution theories and the need to
extend the PN sample currently available are discussed.

\end{abstract}

\begin{keywords}
Stars: abundances -- Stars: AGB and post-AGB. Stars: carbon 
\end{keywords}

\section{Introduction}
The Large Magellanic Cloud, due to its proximity 
\citep[$d \sim$ 50 kpc,][]{feast99} and low average reddening 
\citep[$E(B-V) \sim 0.075$,][]{schlegel98}, has been used successfully as a laboratory to 
test asymptotic giant branch evolution.
The AGB population of the LMC has been thoroughly investigated 
by means of dedicated photometric surveys: the Magellanic Clouds Photometric Survey 
\citep[MCPS,][]{zaritsky04}, the Two Micron All Sky Survey \citep[2MASS,][]{skrutskie06}, 
the Deep Near Infrared Survey of the Southern Sky \citep[DENIS,][]{epchtein94}, Surveying 
the Agents of a Galaxy's Evolution Survey \citep[SAGE--LMC with the {\it Spitzer} 
telescope,][]{meixner06}, and {\it HERschel} Inventory of the Agents of Galaxy Evolution 
\citep[HERITAGE,][]{meixner10, meixner13}. Additional data allowed the reconstruction of the Star 
Formation History (SFH) of the LMC \citep{harris09, weisz13} and the age--metallicity 
relation \citep[AMR,][]{carrera08, piatti13}; these studies confirmed that the stellar
populations are on average sub-solar in metallicity.

Such extensive data sets, interpreted with models for AGB evolution using the metallicities 
of LMC stars, have enlightened and deepened our understanding of the various, still poorly 
known, physical mechanisms characterizing the AGB phase; namely, the third dregde-up, hot 
bottom burning, and the rate of mass-loss. 

The TDU consists in the penetration of the convective envelope following each
thermal pulse. When the bottom of the convective mantle reaches regions of the
star where 3$\alpha$ burning has previously occurred, carbon-rich material is transported to the
surface, with the consequent increase in the surface carbon abundance. Repeated TDU
episodes may lead to the formation of a carbon star, with a surface C/O above unity.
The luminosity function of the carbon star population in the LMC have been extensively
used to draw information on the occurrence of TDU in terms of the extent of the
inward penetration of the surface mantle in the after-pulse phases. The same studies also
allowed  the mass-loss rate experienced by these stars to be calibrated \citep{martin93, 
marigo99, marigo03a, izzard04}.

Hot bottom burning is activated when the temperature at the bottom of the convective 
envelope ($T_{\rm bce}$) reaches sufficiently high values ($T_{\rm bce} > $40 MK) to 
activate proton-capture nucleosynthesis, with the consequent modification of the 
surface chemistry, according to the equilibria of the various reactions involved 
\citep{renzini81, blocker91}. The main effects of HBB are the depletion of the surface 
carbon and nitrogen enrichment. For sufficiently large temperatures 
(above $\sim$90 MK) oxygen destruction also occurs during this process.

In the last few years, stellar evolution models that include the AGB phase have improved, 
with the inclusion of dust production in the stellar wind in the models
\citep{fg06, nanni13a, nanni13b, nanni14, paperI, paperII, paperIII, paperIV}. 
These studies allow a step
forward in our understanding of the physics of these stars because they can be used
to interpret mid-infrared data of the most obscured, dust-enshrouded objects
observed in dedicated surveys such as those mentioned above. The dust surrounding evolved stars
reprocesses the light emitted by the central star at mid-infrared wavelengths. 
The study of the dust formation process is therefore mandatory to interpret their
infrared colours correctly. By treating dust within the models and then comparing the resulting 
yields to observed abundances for similar dust-type stars one can push the comparison 
between yields and observed abundances further than previously done. 
First, by measuring the degree of obscuration of carbon stars through IR observations, 
one can get a handle on the TDU efficiency, since the amount of carbon accumulated at 
the stellar surface, which causes obscuration, depends on the efficiency of the 
TDU mechanism. Second, the amount of silicate dust formed in oxygen-rich stars is tied to
the strength of the HBB experienced. 

`Dusty' AGB models were used by \citet{zhukovska13} and \citet{schneider14} to
calculate the dust production rate by AGB stars in the LMC and compare them with existing 
estimates based on the observations. On the basis of these models, \citet{flavia14, flavia15} 
accomplished a characterization of the obscured stars in the LMC in terms of age, 
metallicity,  and mass distribution. In a recent analysis, \citet{ventura15} showed that LMC 
stars experiencing HBB evolve to well-defined regions of the two-colour infrared diagram. 
Spectroscopic analysis of the selected sample, and in particular measurement of the C/O 
ratio, allows us to determine the strength of the HBB.

In order to gain insight into the LMC AGB stellar population and to get a handle on the
physical mechanisms relevant for their evolution, a good opportunity is offered by the 
study of the planetary nebula population \citep{marigo03b, marigo11, letizia09}. 
Planetary nebulae are related to the final stages of the evolution of AGB stars, 
their ejecta being illuminated by the remnant central star. Their chemical composition 
therefore depends on the relative strength of TDU and HBB experienced during previous stellar phases. 
The chemical composition of PNe, compared to the surface chemistry of AGB stars in their 
final stages, provides insight into the chemical evolution of the star throughout its life, and in 
particular on the mechanisms that alter its surface composition during the final stages 
of its evolution. Whereas the analysis of the distribution of AGB stars in the near- and 
mid-infrared diagnostic diagrams allows a statistical approach to the problem that is useful
for shedding light on the relative duration of the various evolutionary phases, PN abundances 
are strong constraints on stellar surface processes and yields in the final stages of 
AGB stellar life. On the observational side, determination of the surface
chemical composition of PNe is easier than for AGBs because the optical/near-IR 
spectra of such cool giants are contaminated by millions of molecular lines, and deriving 
the abundances of individual species requires the use of spectral synthesis techniques 
\citep[e.g.][]{garcia06, garcia07a, garcia09}, even when taking into account circumstellar
effects \citep{zamora14}. In addition, the more massive and extreme AGBs are
heavily obscured and escape detection (and abundance studies) in the optical
range \citep[e.g.][]{garcia07b} and/or they may display
extremely complex near-IR spectra (e.g.\ McSaveney et al.\ 2007).

Against this background, we embarked on testing the results of suitable AGB stellar evolution yields
\citep{flavia15} to the observed chemical composition of the LMC PN population. The goal of this 
investigation is twofold. On the one hand, we attempt a characterization in terms of the progenitor 
mass and initial chemistry of the LMC PN sample. On the other hand, we use the  PN chemical 
composition to help discriminate among different AGB models and paths, and to clarify 
still open issues related to AGB evolution,
such as the strength of HBB experienced by the different star mass models, the possibility 
that more massive AGB stars may eventually become carbon stars, and the maximum carbon 
enrichment achieved by low-mass AGB stars. We constrain our study to the well-defined LMC 
PN sample in this paper and will extend it  to other PN populations in the future.

In $\S$2 we describe the stellar evolution models. Section 3 describes the changes in surface composition due to 
the AGB processing. 
Section 4 describes the observational sample used in this paper, its limitations and 
uncertainties, and the comparison of PN abundances with the final chemical composition from AGB 
evolution. A discussion is presented in $\S$5, while the conclusions and future 
outlook are in Section 6.

\section{The evolution of AGB stars in the LMC}

Modelling the AGB phase demands a considerable computational effort, owing to the
necessity of adopting extremely short time-scales, as short as a few hours,
during the thermal pulses. 
To date, there are still considerable differences among the results presented by the 
various groups in this field, which just stresses the difficulty in modelling the AGB phase.

The AGB stellar models are extremely sensitive to the input physics, primarily
convection and the rate of mass-loss. In the literature, there are several 
reviews of AGB evolution \citep{herwig05, karakas11, karakas14}, and we do
not repeat them here. In the following subsections we will present a summary 
of the physical input of the AGB models used in this paper.

\subsection{Numerical and physical input}
The AGB models used here are based on the evolution of the 
star and on a description of dust formation in the wind. 

The evolutionary sequences of central stars were calculated by means of the ATON code for 
stellar evolution \citep{mazzitelli89}. The interested reader will find in \citet{ventura98} 
a detailed discussion of the numerical structure of the code; the latest updates are given 
in \citet{ventura09}. Here we briefly recall the most relevant physical input.

The temperature gradient within regions unstable to convective motion was determined
by means of the Full Spectrum of Turbulence \citep[FST,][]{cm91} description. The
efficiency of the convective transport of energy is probably the most important and relevant 
uncertainty affecting the results of AGB modelling \citep{vd05}.

Mass-loss during the AGB for O-rich phases was modelled according to \citet{blocker95}. 
For carbon stars, we used the calibration of $\dot M$, based on hydrodynamical models
of C-star winds, by \citet{wachter02, wachter08}.

In the low-temperature regime (below $10^4$K) we calculated molecular opacities 
by means of the AESOPUS tool, developed by \citet{marigo09}. The advantage of this 
approach is that the opacities are suitably constructed to follow the changes in the 
chemical composition of the envelope driven by TDU and HBB, with the possibility of
accounting for changes in the individual abundances of carbon, nitrogen, and oxygen.
Following this approach is crucial for the description of the carbon-rich
phase because the increase in the molecular opacities occurring when the C/O ratio
approaches (and overcomes) unity favours a considerable expansion of the surface
layers of the star, with the consequent enhancement of the rate at which mass loss occurs
\citep{vm10}.

The dust formation process is described in \citet{flavia15}.
To span the range of metallicities of stars in the LMC \citep{harris09}, we used three 
sets of models with metallicity $Z=10^{-3}$, $4\times 10^{-3}$, and $8\times 10^{-3}$.

Each model was evolved starting from the pre--main sequence phase until when almost all
the external envelope was lost; prosecuting the computations until the beginning of the
PN phase would demand a much greater computational effort, with no significant
improvement to the results needed in this context.

\subsection{Physical properties of AGB evolution models} 
\label{agbmodel}
The models presented here, including the discussion of the evolutionary sequences, 
are extensively illustrated in \citet{ventura14b} ($Z=4\times 10^{-3}$), \citet{ventura13}
($Z=1,8\times 10^{-3}$, initial mass above $3~M_{\odot}$) and \citet{paperIV} (low--mass
models of metallicity $Z=1,8\times 10^{-3}$ and initial mass below $3~M_{\odot}$).
Here, we review the relevant aspects for the present paper regarding these models.

High-mass stellar models evolve to more massive AGB stellar cores, 
independently of the progenitor metallicity. These massive cores correspond to a higher 
stellar luminosity, hence, given our mass-loss prescription, to a higher rate of mass-loss. 
Stars with initial masses in the range $1.25~M_{\odot} \leq M_{\rm i} \leq 3~M_{\odot}$
reach the C--star stage after a series of thermal pulses, each associated with a TDU 
episode, that gradually increases the surface stellar carbon. These limits are slightly 
dependent on metallicity: the lower limit to reach the C--star stage is $1~M_{\odot}$ for
$Z=1,4\times 10^{-3}$, whereas it is $1.25~M_{\odot}$ for $Z=8\times 10^{-3}$; the upper 
limit in the initial mass to enter the C--star phase is $3~M_{\odot}$ for
$Z=4,8\times 10^{-3}$, whereas it is $2.5~M_{\odot}$ for $Z=10^{-3}$.

After reaching a surface abundance ratio of $C/O>1$, the surface molecular opacity 
increases \citep{marigo02, vm09, vm10}, thereby triggering expansion (and cooling) of the external 
regions. As a consequence, the rate of envelope mass loss is enhanced when C/O becomes
greater than unity.

A clear separation distinguishes models with initial mass below $\sim 3~M_{\odot}$ from their
higher mass counterparts, which experience HBB.  
Stars that go through HBB will never become carbon stars, because their surface carbon is rapidly 
destroyed by proton-capture nucleosynthesis occurring at the base of the convective 
envelope.

The trend with mass in the high-mass domain (M$>3~M_{\odot}$) is straightforward: the 
higher the initial stellar mass, the higher is the temperature $T_{\rm bce}$ at which HBB 
occurs, the steeper the core mass vs.\ luminosity relationship. These high-mass models 
lose the convective envelope very rapidly and thus experience a limited number of thermal 
pulses \citep[see Table 1 in][]{ventura13}, which prevents the possibility of any 
contamination of the surface chemistry by TDU.

A word of caution is needed regarding the dependence of models on physical
input, particularly on convection and mass-loss rate. Models calculated with
a less efficient treatment of convection than used here, such as the traditional mixing 
length (ML) scheme, experience weaker HBB, thus evolving at lower luminosities and
losing their envelopes at a lower rate. If the ML scheme were applied, the number of
thermal pulses experienced by massive AGB models would be higher than the number 
of pulses suffered by low-stellar mass models. As a result, massive stars would evolve as 
carbon stars in the very final AGB phases. An exhaustive discussion of this argument can 
be found in the detailed analysis by \citet{vd05} and in the more recent investigations 
by \citet{doherty14a, doherty14b}.

\begin{figure*}
\begin{minipage}{0.33\textwidth}
\resizebox{1.\hsize}{!}{\includegraphics{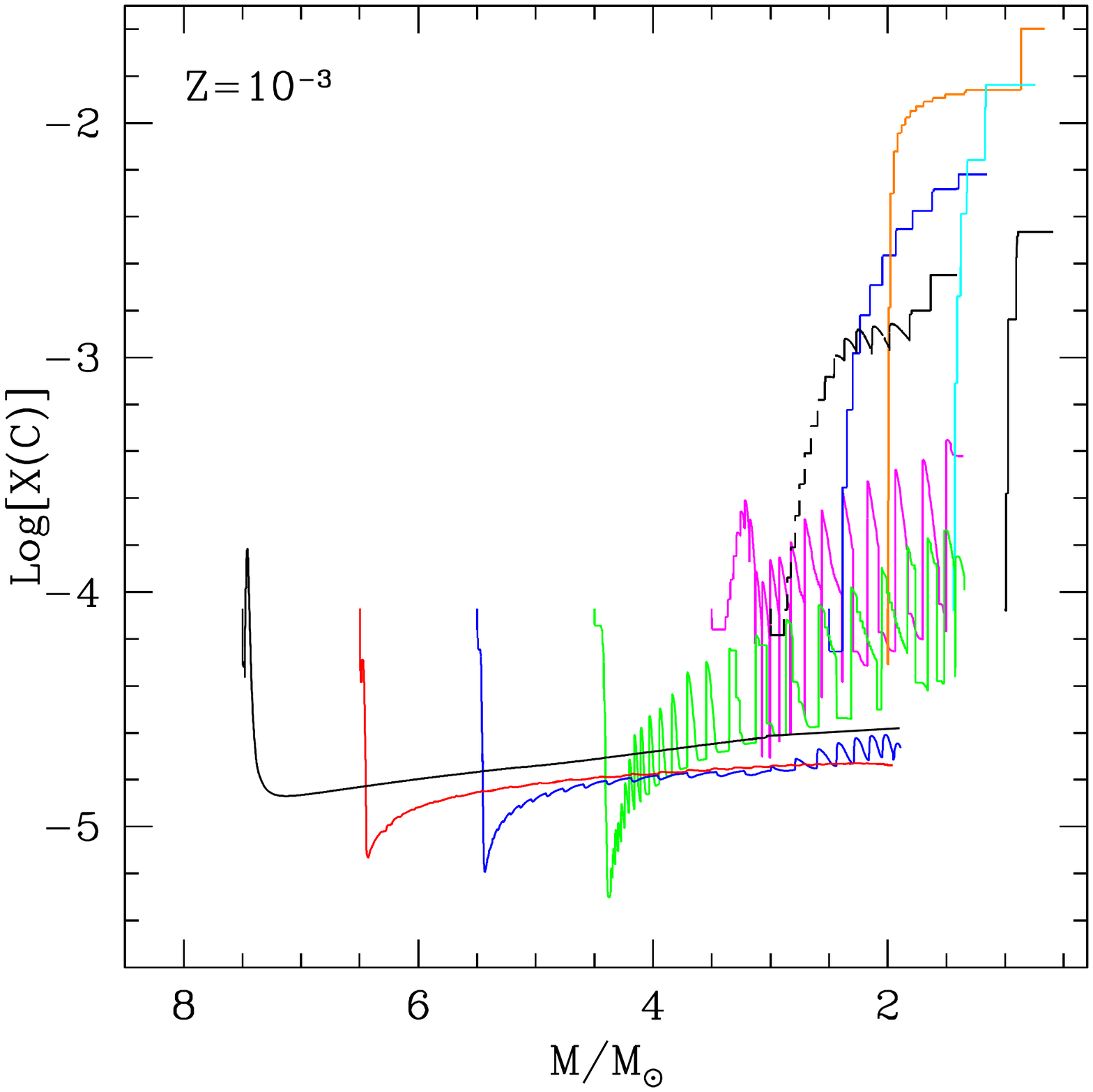}}
\end{minipage}
\begin{minipage}{0.33\textwidth}
\resizebox{1.\hsize}{!}{\includegraphics{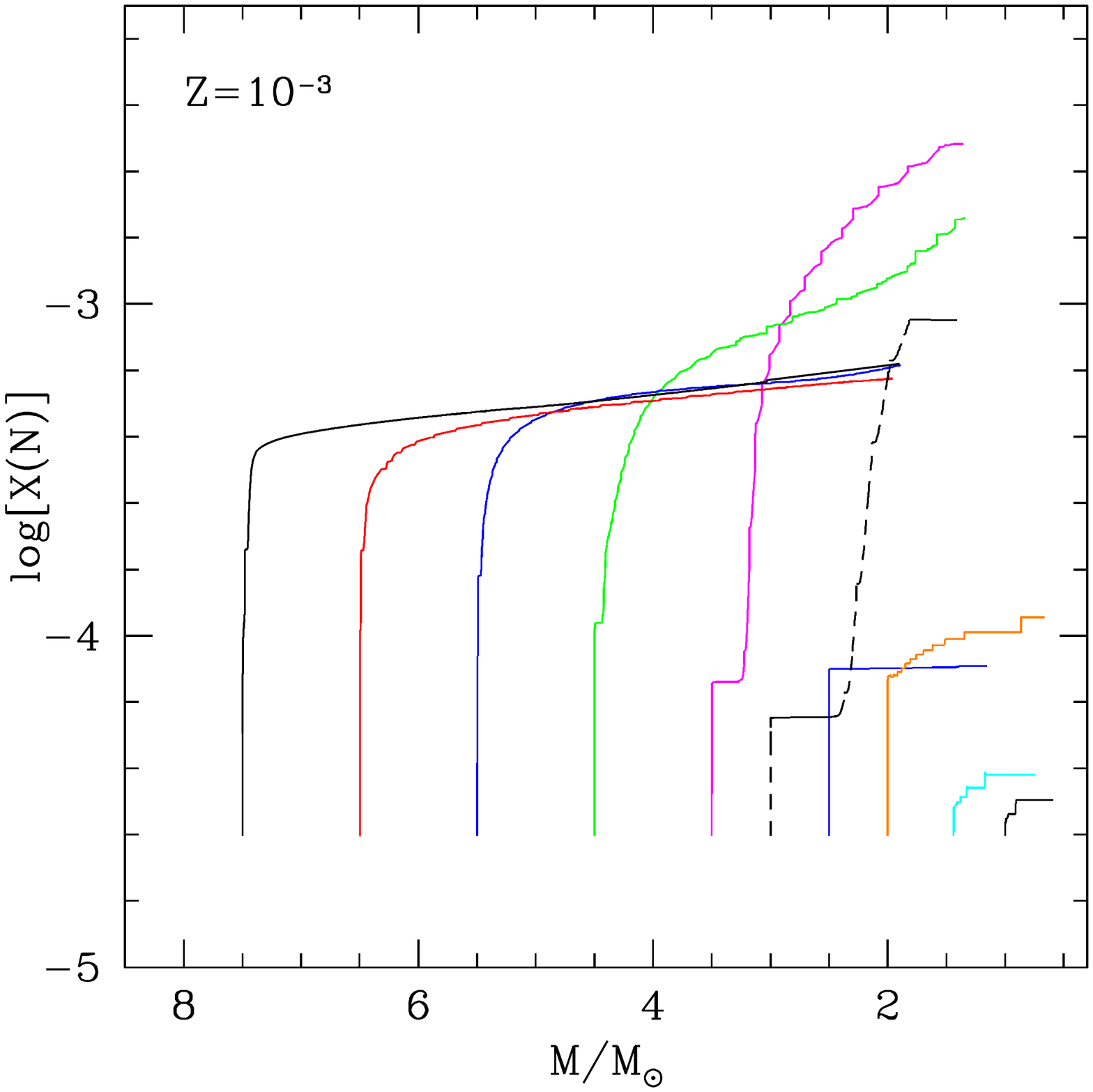}}
\end{minipage}
\begin{minipage}{0.33\textwidth}
\resizebox{1.\hsize}{!}{\includegraphics{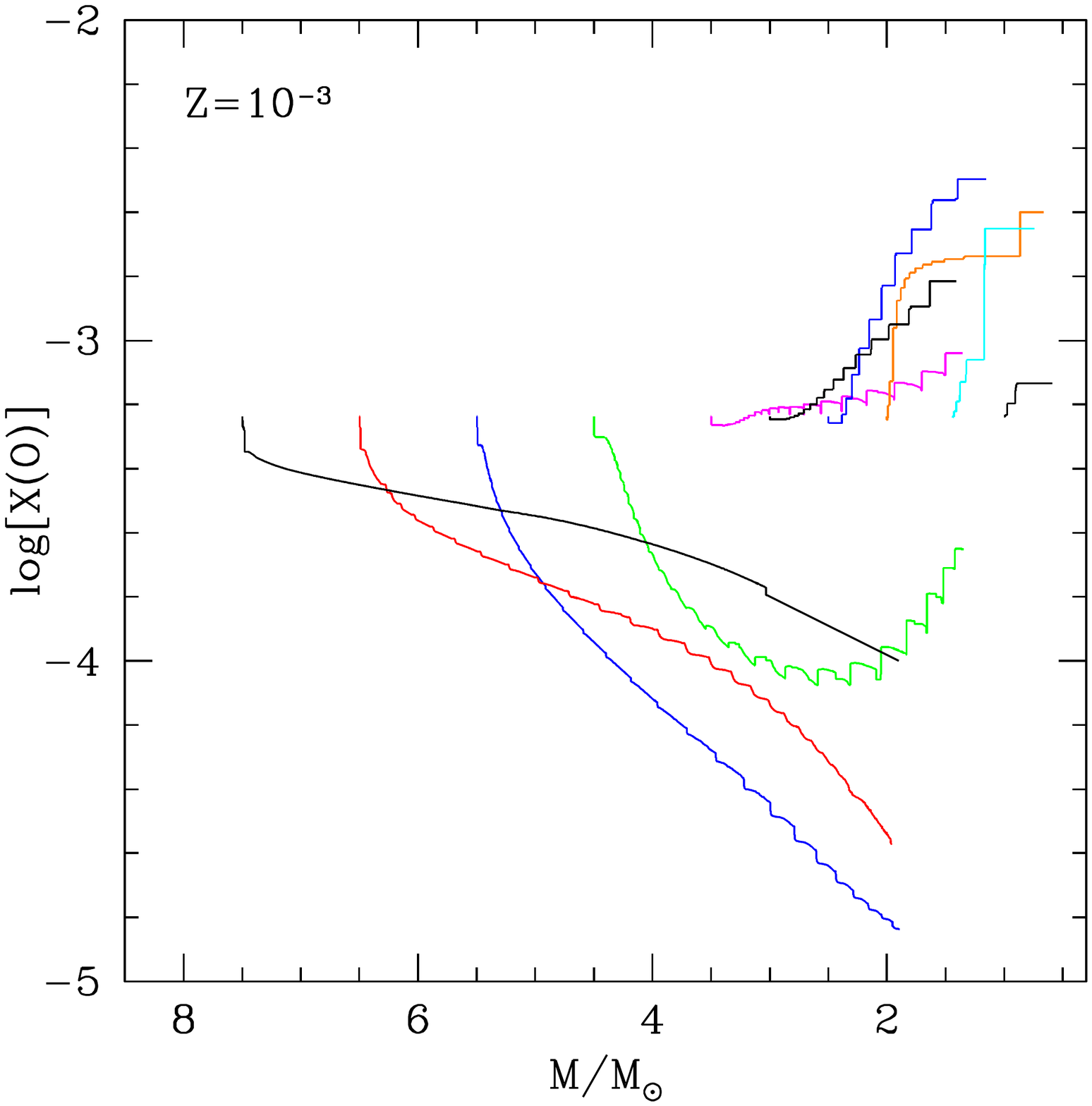}}
\end{minipage}
\vskip-30pt
\caption{The evolution of the surface mass fraction of carbon (left), nitrogen 
(middle) and oxygen (right) during the AGB phase of models of initial mass in the range
$1~M_{\odot} \leq M \leq 7.5~M_{\odot}$ and metallicity $Z=10^{-3}$. Along the abscissa we 
report the total mass of the star (decreasing during the evolution). For clarity, we
show only the evolution of models of initial mass $1$, $1.5$, $2$, $2.5$, $3$, $3.5$, $4.5$,
$5.5$, $6.5$, and $7.5M_{\odot}$.
}
\label{fz1m3}
\end{figure*}

\begin{figure*}
\begin{minipage}{0.33\textwidth}
\resizebox{1.\hsize}{!}{\includegraphics{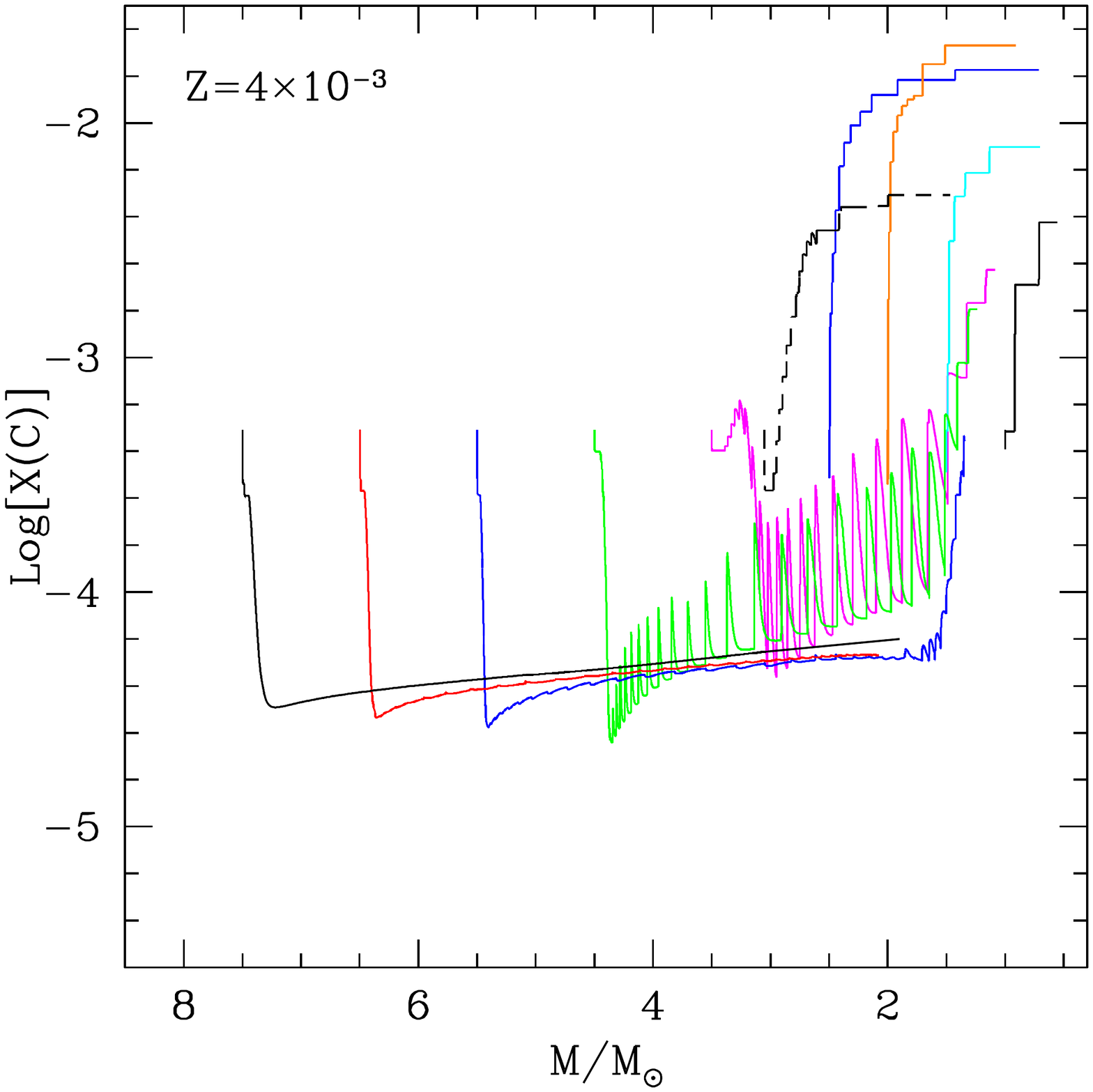}}
\end{minipage}
\begin{minipage}{0.33\textwidth}
\resizebox{1.\hsize}{!}{\includegraphics{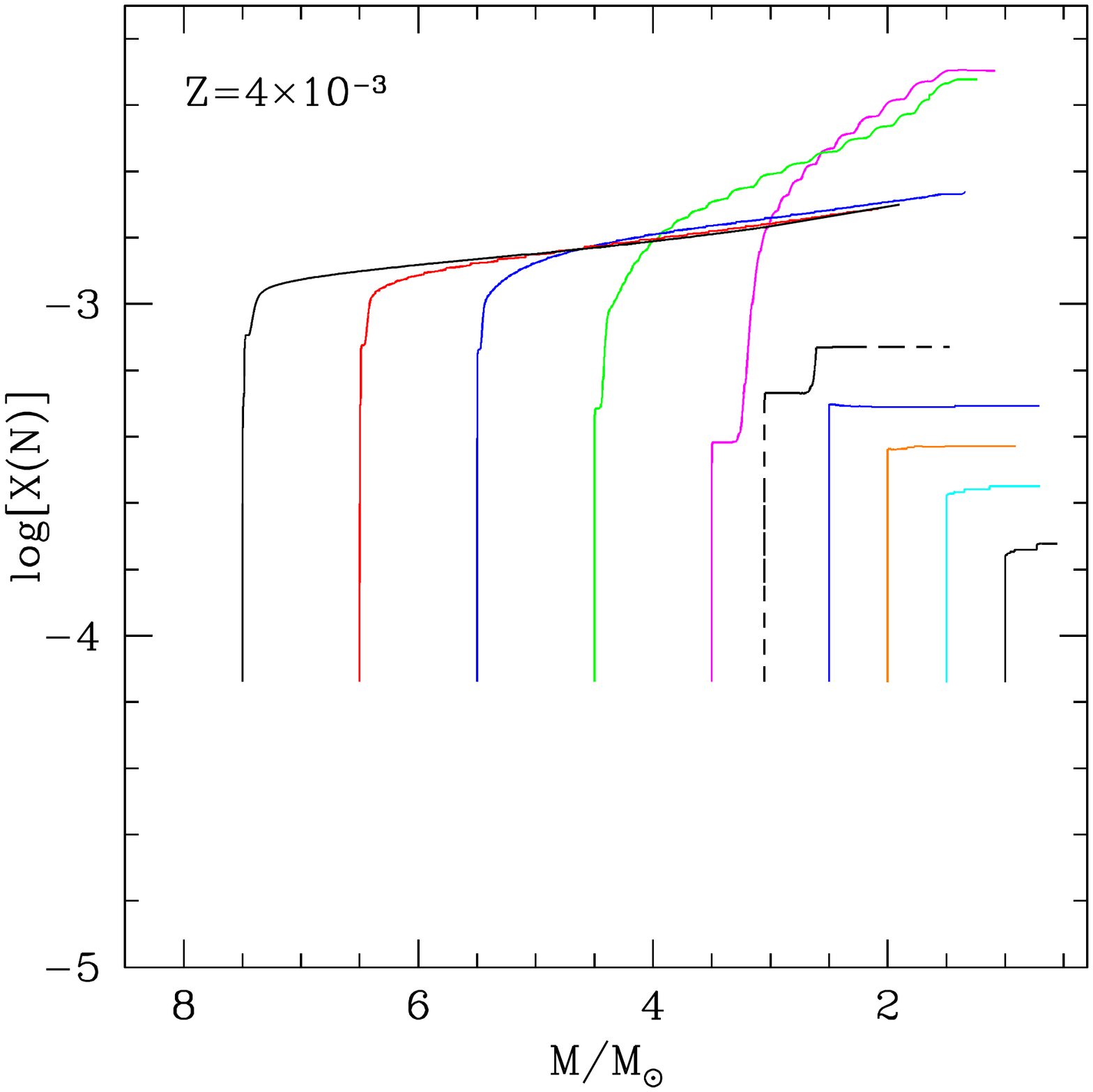}}
\end{minipage}
\begin{minipage}{0.33\textwidth}
\resizebox{1.\hsize}{!}{\includegraphics{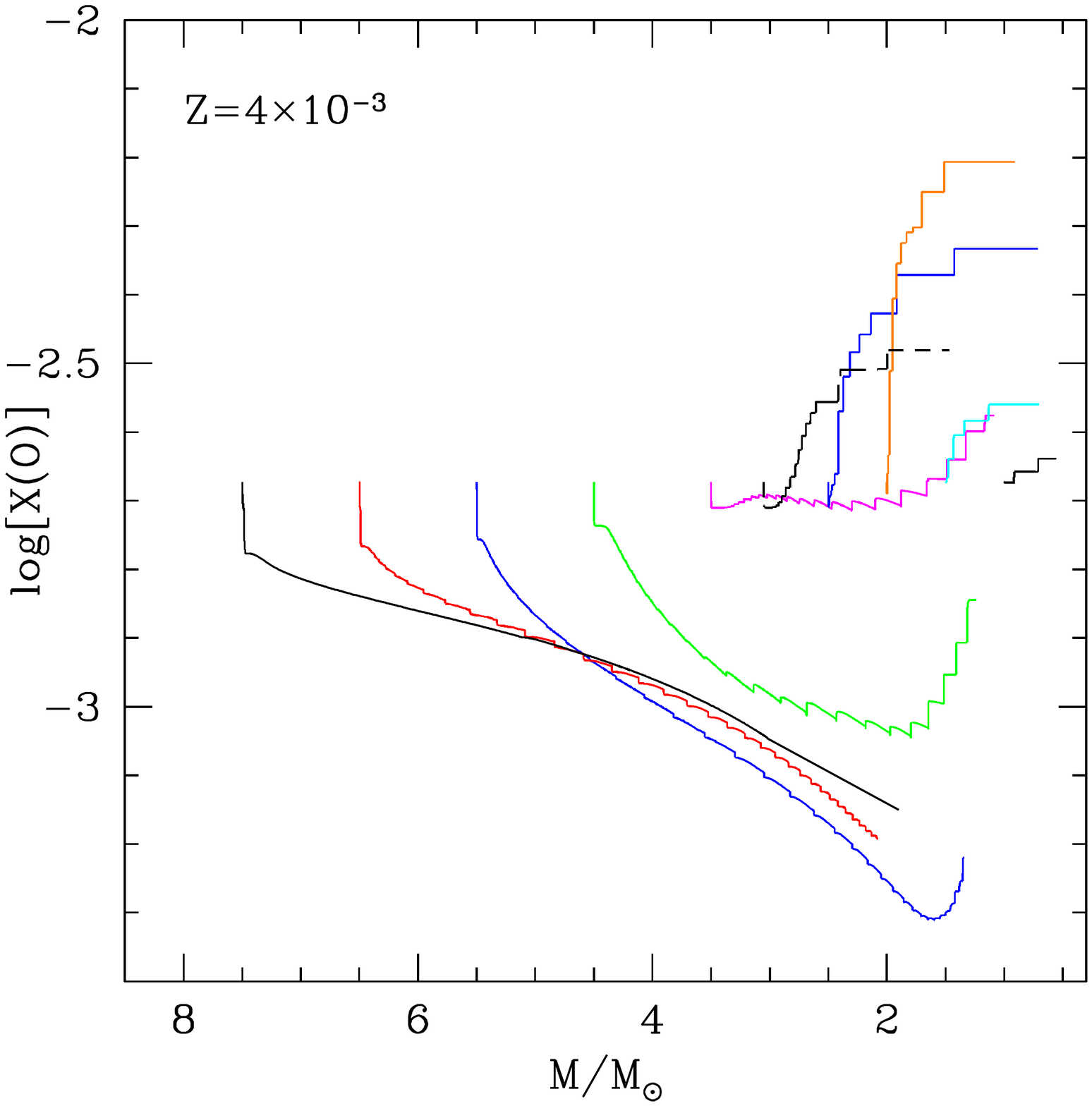}}
\end{minipage}
\vskip-30pt
\caption{The same as in Fig.\ \ref{fz1m3}, but referred to AGB models of metallicity
$Z=4\times 10^{-3}$.}
\label{fz4m3}
\end{figure*}

\begin{figure*}
\begin{minipage}{0.33\textwidth}
\resizebox{1.\hsize}{!}{\includegraphics{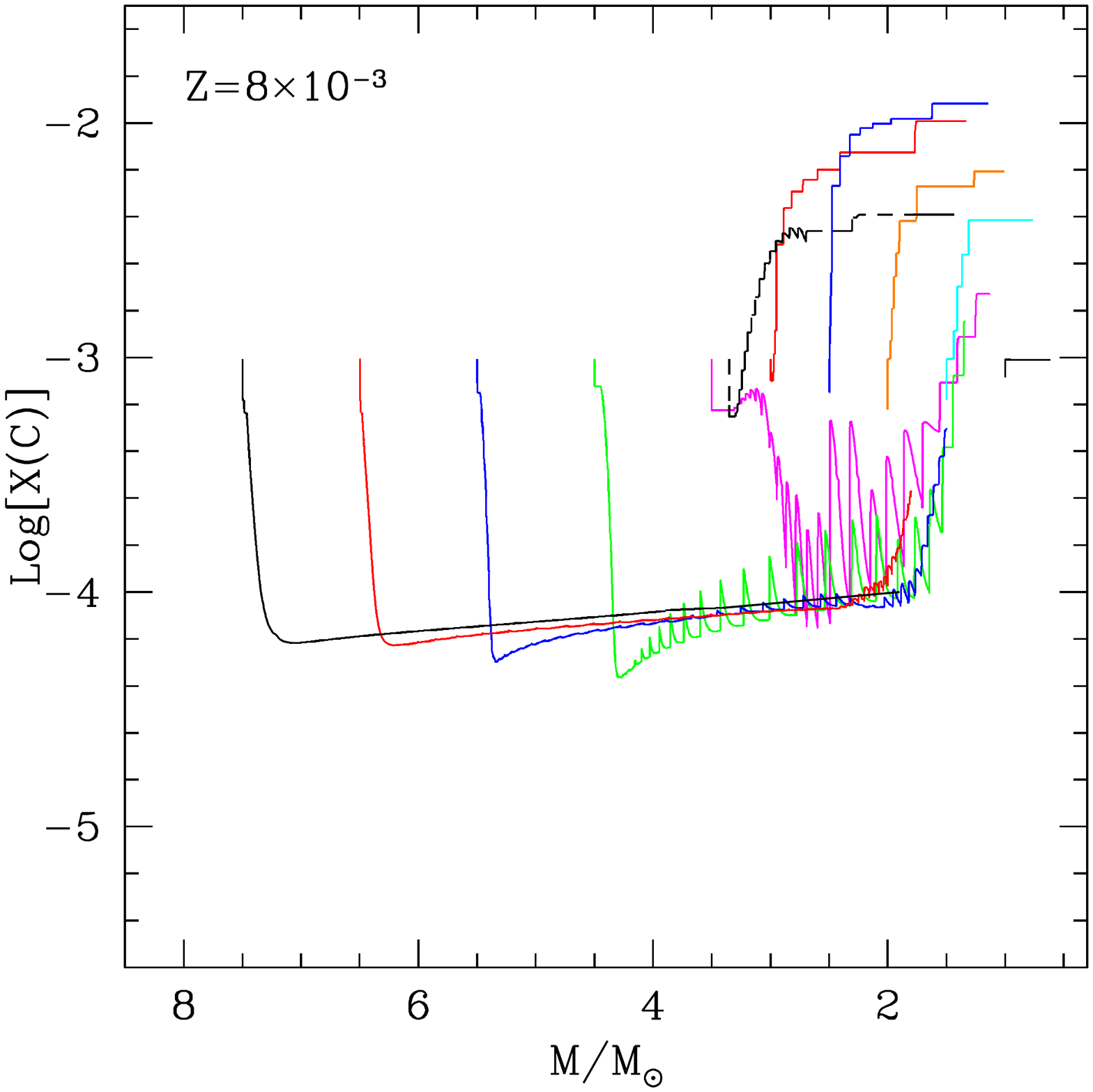}}
\end{minipage}
\begin{minipage}{0.33\textwidth}
\resizebox{1.\hsize}{!}{\includegraphics{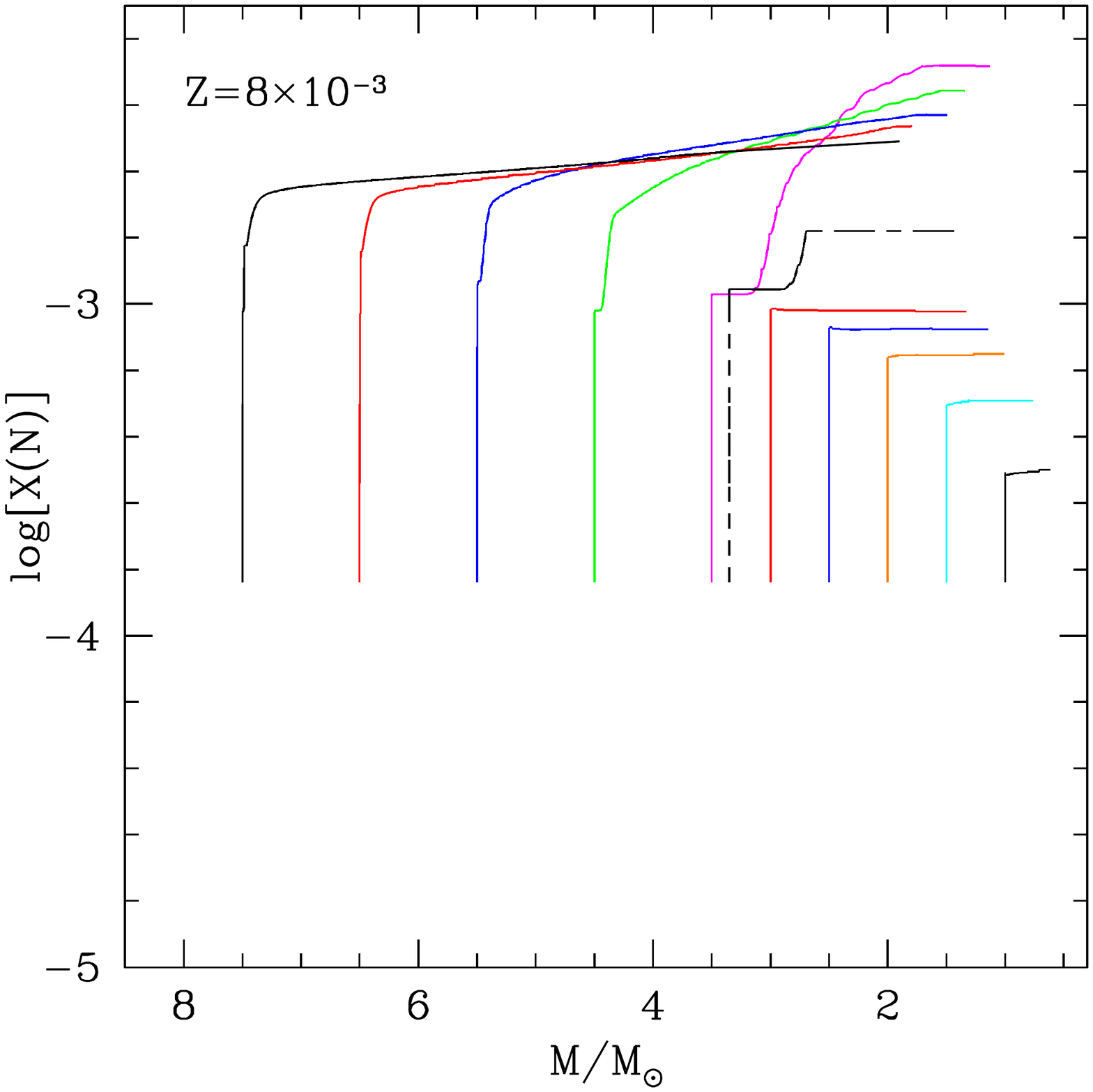}}
\end{minipage}
\begin{minipage}{0.33\textwidth}
\resizebox{1.\hsize}{!}{\includegraphics{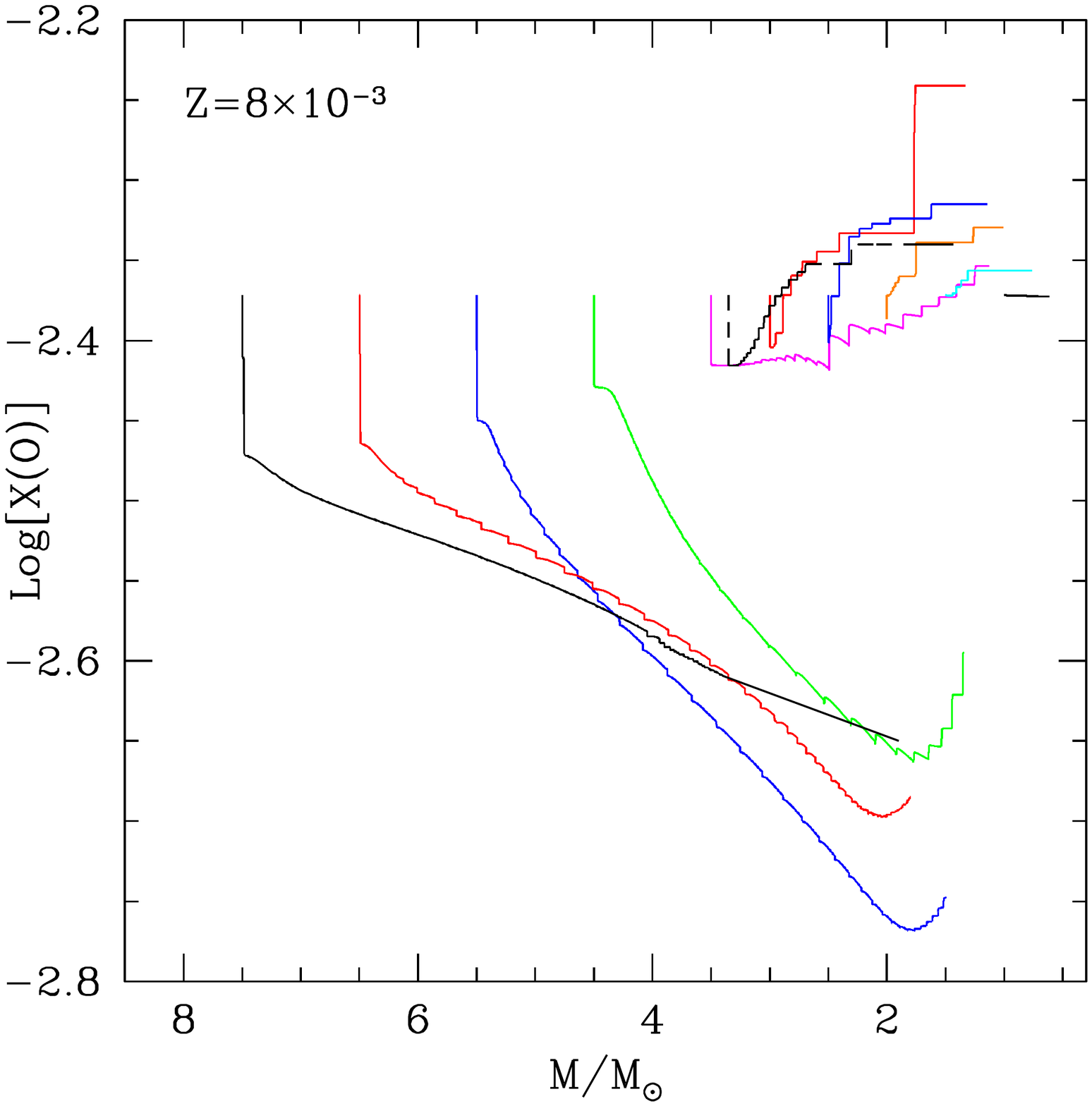}}
\end{minipage}
\vskip-30pt
\caption{The same as in Fig.\ \ref{fz1m3}, but referred to AGB models of metallicity
$Z=8\times 10^{-3}$.}
\label{fz8m3}
\end{figure*}

\section{Change in the surface chemistry of AGB stars}

The variation of the surface chemical composition of the AGB models in this paper, in 
terms of the surface mass fractions of the CNO elements, is shown in Fig.\ \ref{fz1m3}, 
\ref{fz4m3}, and \ref{fz8m3}, respectively for $Z=10^{-3}$, $Z=4\times 10^{-3}$, and 
$Z=8\times 10^{-3}$. The abscissae give total stellar mass. This choice allows us to show 
the sequences of the different masses in the same plane, despite the difference in the 
evolutionary times. The initial mass of each model can be deduced from the abscissa in
the starting point of each track.

\subsection{Low-mass AGB stars: the effect of Third Dredge--Up}
\label{lowmass}
The variation of the surface carbon, shown in the left panels of Fig.\ \ref{fz1m3}, 
\ref{fz4m3}, and \ref{fz8m3}, reflects the relative contributions from TDU and HBB
in modifying the surface chemical composition. In low-mass stars ($M<3~M_{\odot}$), 
TDU is the only active mechanism; thus, the surface carbon increases during AGB 
evolution. For all the metallicities investigated, the largest abundances of carbon,
$X(C) \sim 10^{-2}$, are reached by models with mass $M \sim 2$--$2.5~M_{\odot}$. The overall 
surface carbon abundance increases by a factor of $\sim$$10$ for $Z=8\times 10^{-3}$, 
and by two orders of magnitude for $Z=10^{-3}$. Models with mass below $\sim$$2~M_{\odot}$ 
experience a smaller number of thermal pulses than the high-mass ones; hence, their
final carbon abundance is smaller. Note that the final carbon mass fraction for stars experiencing TDU 
is practically independent of the initial abundance; rather, it is 
essentially determined by the extent of the inward penetration of the convective envelope 
into regions previously touched by $3\alpha$ burning during each TDU episode.

Stars of initial mass below $\sim 1.25~M_{\odot}$ (but see discussion in section \ref{agbmodel} on 
the sensitivity of this limit to the metallicity) do not reach the final carbon star phase because 
they lose the external mantle after a few thermal pulses, thus experiencing only a 
limited number of TDU episodes. In fact, the surface chemistry of these stars is changed 
only by the first dredge-up, which occurs while ascending the red giant branch (RGB).

The surface oxygen also increases in low--mass AGBs (see right panels of Fig.\ \ref{fz1m3}, 
\ref{fz4m3}, and \ref{fz8m3}) because, during TDU, the base of the surface
convection reaches layers where some production of oxygen has occurred. The percentage 
variation in the surface oxygen is smaller (by a factor of $\sim$2--3) with respect to 
carbon. The final oxygen surface abundance is fairly independent of the initial stellar 
abundance.

Unlike carbon, the surface nitrogen is not expected to undergo significant changes in 
low--mass AGB stars, which renders their final nitrogen abundances sensitive to the assumed 
initial relative abundances and to the metallicity of the star.

\subsection{Massive AGB stars: the signature of hot bottom burning}
\label{hbbmodel}

During the AGB evolution of the higher-mass stars, those that undergo HBB, the variation 
of the surface chemical composition depends on the 
strength of the HBB experienced. The effects of HBB can be seen in the sudden decrease 
(by a factor of $\sim$20) of surface carbon, occurring in the initial phases of 
AGB evolution for $M \geq 4~M_{\odot}$ (see left panels of Fig.\ \ref{fz1m3}, 
\ref{fz4m3}, and \ref{fz8m3}). The depletion of surface carbon occurs in conjunction
with the increase in nitrogen abundance (see middle panels of the figures).
The final abundances of C and N depend on whether the models  suffer TDU in the final 
evolutionary phases.

Models with mass near the upper limit of AGB evolution ($M \sim 6$--$8~M_{\odot}$) are 
expected to experience only a small number of weak thermal pulses.
The chemistry of these models is not contaminated by TDU, so their
final abundances will reflect the pure effects of HBB. Consequently, they will end
their AGB history with a carbon abundance a factor  $\sim$10 lower than the
initial abundance. Nitrogen will be greatly enhanced (by $\sim$20--30) owing to the combined 
effects of HBB and of the first dredge-up.
The final abundances of carbon and nitrogen in these stars will be determined by the
equilibra of proton-capture nucleosynthesis and will therefore scale with the initial
overall C+N+O content of the star.

Models of mass $4~M_{\odot} \leq M \leq 6~M_{\odot}$ undergo a more complex evolution,
because the initial contamination by HBB is followed by surface carbon enrichment,
caused by TDU. This can be seen in the left panels of Fig.\ \ref{fz1m3}, \ref{fz4m3}, 
and \ref{fz8m3}, where the surface carbon undergoes a series of ups and downs,
the signature of the combined effects of TDU and HBB. In the very final AGB phases
the mass of the envelope falls below the threshold for HBB: the surface
carbon is determined solely by TDU so that it gradually increases until the end of the
AGB evolutionary phase. 
The final carbon abundance depends on the number of TDU episodes 
experienced in the very final phases, when HBB is switched off. The present computations 
indicate that the final carbon in the star is higher the closer its initial mass is to
the lower limit to activate HBB (i.e,\ to $\sim$3.5--$4~M_{\odot}$).
Nitrogen increases compared to its initial value, even more so than in the
most massive models, owing to the bounty of carbon available both originally and from 
subsequent dredge-ups from the ashes of helium-burning. 

\subsection{The minimum threshold mass to activate HBB}
\label{middle}
Models of mass close to the threshold mass for HBB activation, $M \sim 3~M_{\odot}$,
are shown with dashed lines in Fig. \ref{fz1m3}, \ref{fz4m3}, and \ref{fz8m3}; they
exhibit an interesting and specific behaviour. These stars experience HBB until the 
repeated TDU episodes lead to the formation of a carbon star; when the C-star stage is 
reached HBB is extinguished, owing to the cooling of the external
regions, favoured by the increase in the molecular opacities \citep{marigo07, vm09}. 
Note that this behaviour is found only when the low-temperature, molecular opacities 
for C-rich gas are used \citep{marigo02}; in models using opacities calculated by 
neglecting carbon enhancement, HBB remains efficient until the whole envelope is lost.
 
The final chemistry of these stars will be somewhat intermediate between low-mass AGB 
stars and more massive stars experiencing full HBB: the surface N will be slightly 
enhanced compared to the original abundance, whereas their surface carbon will be much 
higher than at the beginning of their evolution.

\subsection{The puzzling behaviour of oxygen}
\label{oxygen}
Oxygen deserves separate discussion in this context. As shown in the right panels of
Fig. \ref{fz1m3}, \ref{fz4m3}, and \ref{fz8m3}, the surface oxygen decreases in all
models experiencing HBB, the signature of the activation of CNO cycling. Whereas
the CN cycle is efficiently activated in all cases, because it requires temperatures of the
order of $\sim$40 MK, full CNO burning, with the destruction of the surface oxygen, 
requires temperatures $T \sim$ 90 MK, which renders the results extremely sensitive
to metallicity. Because low-$Z$ models experience more efficient HBB \citep{ventura13}, 
the depletion of the surface oxygen with respect to the original content is much higher
in the $Z=10^{-3}$ models ($\delta log[X(O)] \sim 0.7$--1.5), compared to $Z=4\times 10^{-3}$ 
($\delta {\rm log}[X(O)] \sim 0.5$--0.8) and $Z=8\times 10^{-3}$ ($\delta {\rm log}[X(O)] \sim 0.2$--0.3). 

The trend of oxygen abundances with mass is not trivial: the stars showing the most extreme chemistry
are those with mass $\sim$5--6~$M_{\odot}$, which end their evolution with less
oxygen compared to their higher-mass counterparts. The reason for this is that
models of higher mass lose the convective envelope very rapidly, before a very
advanced nucleosynthesis has occurred. This effect, as discussed in details in
\citet{ventura13}, is a consequence of the use of the FST model for convection and of
the \citet{blocker95} treatment of mass loss.

Although the evolutionary sequences used in the present investigation are interrupted 
before the total consumption of the external mantle, the computations are extended to
a sufficiently advanced stage during the AGB evolution that the final surface chemical
composition of the models can be directly compared with the observed chemical abundances
of PNe. Low--mass AGBs suffer a strong mass loss after becoming carbon stars, owing to the
effects of radiation pressure on solid carbon grains formed in the wind of the star. The
loss of the envelope becomes faster and faster as more carbon is accumulated to the
surface layers. In all the low--mass models discussed here the calculations reach phases
when the loss of the envelope became so fast that the possibility that additional
TDU episodes modify the surface chemical composition can be disregarded. On the side of 
massive AGBs, the nuclear activity at the base of the envelope is progressively extinguished by the loss
of the external mantle; this is due to the general cooling of the external regions close
to the bottom of the envelope. The computations of models in this range of masses were
extended in all cases until the HBB was almost completely extinguished. Had we followed
the evolution until the full ejection of the envelope, we would find a slightly higher
nitrogen content and a smaller oxygen abundance; however, the differences would be
significantly smaller than the errors associated to the observations, presented in the
next section.

\section{Planetary nebulae in the LMC}

\subsection{The LMC PN database}

In order to compare data and models we have used a homogeneous sample of LMC PNe whose abundances have
been collected over the years based on ground and space data sets. The most important 
element for these comparisons, carbon, has been observed directly with STIS/{\it HST} in 
22 PNe \citep{letizia05}. Another 17 PNe have reliable carbon 
abundances available in the literature (Leisy \& Dennefeld 2006); upper limits to carbon 
for four additional PNe, and uncertain carbon determination for three PNe, are also given in the 
latter reference, which includes other critical elemental abundances such as helium, 
nitrogen, and oxygen. 

In Table 1 we give the database used in this study. Column (1) gives the most used PN name. 
Column (2)  gives the published morphological class. Columns (3) through (5) give 
respectively the gas-phase abundances of C, N, and O (by number) of the PNe, in the usual 
format log($X$/H)+12. Finally, column (6) gives, where available, the dust type from 
\textit{Spitzer}/IRS observations \citep{letizia07}. We list in 
this table all LMC PNe that have reliable measured abundances for at least one of the CNO 
elements. All abundances in Table~1 are from Leisy \& Dennefeld (2006) and 
references therein. All abundances in Table 1 have been calculated with the direct method, 
i.e.\ through the calculation 
of electron temperature and density, and with unobserved ion emission accounted for via 
the ionization correction factor (ICF) method (Kingsburgh \& Barlow 1994). Note that the 
flux and reddening correction uncertainties for these samples are minimal when compared to 
the uncertainties produced by the ICF method. Most PNe in which carbon has been detected are 
in such an excitation range that the 
observation of C$^{\rm +}$, C$^{\rm 2+}$, and C$^{\rm 3+}$ directly delivers the total 
carbon abundance. The derived uncertainties in carbon abundances are thus small, with both 
\citet{letizia05} and Leisy \& Dennefeld (2006) 
indicating uncertainties  $<$0.1 dex in 12+log(C/H). Uncertainties in the other elements 
are 0.01 dex for helium, 0.1 dex in oxygen and neon, and 0.15-0.2 dex in nitrogen 
abundances (Leisy \& Dennefeld 2006). For the three unreliable abundances mentioned above, 
the uncertainty can be up to 0.3--0.5 dex.

Nebular morphology is available for most the PNe listed in the table. They are taken directly from 
\citet{letizia00} and  \citet{shaw01, shaw06}. Here, we list only the main shapes, 
i.e.\ round (R), elliptical (E), bipolar or quadrupolar (B or Q), bipolar core (BC), and 
point-symmetric (P). In some cases the morphology is uncertain, as noted 
in the table.

Another physical characteristic that we need for this study is the dust content of the 
PNe. \citet{letizia07} have observed a sizeable sample (25)
of LMC PNe with Spitzer/IRS spectroscopy, finding different dust compositions
(dust types) in their circumstellar envelopes. About half of the total sample
shows carbon-rich dust features (nine C-rich PNe) or featureless spectra (14 F PNe,
dust-free and quite evolved PNe dominated by nebular emission lines), while a
small minority (two O-rich PNe) show oxygen-rich dust features.

\setlength{\tabcolsep}{0.1cm}
\begin{table}
\caption{}
\begin{tabular}{l l l l l l}
\hline 
Name & M& (C/H) & (N/H) & (O/H) & dust type\\
\hline 
 MG~45 &      E & $\dots$ &  8.41     &  8.27 & $\dots$ \\
SMP~01 &      R &  8.40   & $\dots$   &  8.34 & $\dots$ \\
SMP~02 &      R & $\dots$ &  6.95     &  8.03 & $\dots$ \\
SMP~03 &      R & $\dots$ &  7.03     &  7.75 & $\dots$ \\
SMP~04 &      E &  8.66\textsuperscript{a}    &  7.93    &  8.61   & F \\
SMP~05 &      E & $\dots$ &  6.34     &  8.00 & $\dots$ \\
SMP~08 &      R & $\dots$ &  8.00     &  8.16 & $\dots$ \\
SMP~09 &     BC &  8.43\textsuperscript{a}    & $\dots$  & $\dots$ & C-rich \\
SMP~10 &      P &  7.35\textsuperscript{a}    & $\dots$  & $\dots$ & F \\
SMP~11 &      B & $\dots$ &  7.10     &  7.18 & $\dots$ \\
SMP~13 &     BC &  7.92   & $\dots$   &  8.39 & $\dots$ \\
SMP~16 &      B &  7.33\textsuperscript{a}    &  8.65    &  8.32   & F \\
SMP~18 &     BC\textsuperscript{b}   &  8.37\textsuperscript{a}    & $\dots$   & $\dots$ & F \\
SMP~19 &     BC &  8.46\textsuperscript{a}    & $\dots$  & $\dots$ & C-rich \\
SMP~21 &      Q &  7.34   &  8.37    &  7.86  & O-rich \\
SMP~25 &      R &  8.29\textsuperscript{a}    & 7.26     &  8.17   & C-rich \\
SMP~27 &      Q &  7.84\textsuperscript{a}    &  7.10    &  8.31   & F \\
SMP~28 &      P &  8.01   & $\dots$  &$\dots$ & $\dots$ \\
SMP~29 &      B &  7.49   &  8.79    &  8.05  & $\dots$ \\
SMP~30 &      B\textsuperscript{b}   &  7.75\textsuperscript{a}    &  8.55     &  8.24 & $\dots$ \\
SMP~31 &      R\textsuperscript{b}   &$\dots$ &  7.03    &  7.29   & $\dots$ \\
SMP~32 &      R & $\dots$ &  7.73    &  8.39  & $\dots$ \\
SMP~33 &     BC & $\dots$ &  7.84    &  8.91  & $\dots$ \\
SMP~34 &      E &  8.13\textsuperscript{a}    & $\dots$  & $\dots$ & F \\
SMP~40 &     ES & $\dots$ &  8.14    &  8.56  & $\dots$ \\
SMP~45 &      B\textsuperscript{b}   &  7.58\textsuperscript{a}    & $\dots$   & $\dots$ & F \\
SMP~46 &     BC &  8.76\textsuperscript{a}    & $\dots$  & $\dots$ & C-rich \\
SMP~47 &      E &  8.56 &  8.69      &  8.25  & $\dots$ \\
SMP~48 &      E &  8.40\textsuperscript{a}    & $\dots$  &  8.24   & C-rich \\
SMP~49 &      R & $\dots$ &  7.60    &  8.33  & $\dots$ \\
SMP~53 &     BC & $<$6.70 &  7.72    &  8.23  & $\dots$ \\
SMP~54 &      B & $<$8.14 & $\dots$  &$\dots$ & $\dots$ \\
SMP~55 &      R & $\dots$ &  7.27    &  8.40  & $\dots$ \\
SMP~59 &      B &  7.16\textsuperscript{a}    &  8.49    &  8.52   & $\dots$ \\
SMP~62 &     BC &  7.27   & $\dots$  &$\dots$ & $\dots$ \\
SMP~63 &      E &  8.80   & $\dots$  &  8.39  & $\dots$ \\
SMP~66 &      E &  8.51   &  7.61    &  8.31  & F \\
SMP~67 &      B & $<$7.66 &  8.00    &  8.50  & $\dots$ \\
SMP~68 &      E & $\dots$ & $\dots$  &  8.94  & $\dots$ \\
SMP~69 &      B & $<$8.32 &  8.61    &  8.63  & $\dots$ \\
SMP~71 &      E &  8.90\textsuperscript{a}    &  8.03    &  8.63   & C-rich \\
SMP~72 &      B &  8.21\textsuperscript{a}    & $\dots$  & $\dots$ & F \\
SMP~73 &     BC &  8.78   & $\dots$  &  8.66  & $\dots$ \\
SMP~75 &      R & $\dots$ &  7.44    &  8.29  & $\dots$ \\
SMP~77 &      E &  9.12   & $\dots$  &  8.27  & $\dots$ \\
SMP~78 &     BC &  8.51   & $\dots$  &$\dots$ & $\dots$ \\
SMP~79 &     BC &  8.67\textsuperscript{a}    &  8.02    &  8.34   & C-rich \\
SMP~80 &      R &  7.51\textsuperscript{a}    &  7.39    &  8.34   & F \\
SMP~81 &      R\textsuperscript{b}   &  7.16\textsuperscript{a}    &  7.12     &  8.25 & O-rich \\
SMP~82 &      E & $\dots$ &  8.59    &  8.16  & $\dots$ \\
SMP~83 &      B &  7.49   & $\dots$  &$\dots$ & $\dots$ \\
SMP~84 &      E & $\dots$ &  7.43    &  8.15  & $\dots$ \\
SMP~85 &      R &  8.74   &  7.26    &  8.00  & $\dots$ \\
SMP~87 &     BC & $\dots$ &  8.74    &  8.23  & $\dots$ \\
SMP~88 &      E & $<$8.84 &  7.79    &  7.91  & $\dots$ \\
SMP~89 &     BC\textsuperscript{b}   &$\dots$ & $\dots$  & $\dots$ & $\dots$ \\
SMP~91 &      B & $\dots$ &  8.48    &  8.27  & $\dots$ \\
SMP~92 &     BC &  8.21   &  7.77    &  8.62  & $\dots$ \\
SMP~93 &      B &  7.62\textsuperscript{a}    &  8.69    &  8.58   & $\dots$ \\
SMP~95 &     BC &  8.82\textsuperscript{a}    & $\dots$  & $\dots$ & F \\
SMP~96 &     BC & $\dots$ &  8.32    &  8.04  & $\dots$ \\
SMP~97 &      R &  8.50   & $\dots$  &  8.50  & F \\
SMP~98 &      R & $\dots$ & $\dots$  &  8.57  & $\dots$ \\
SMP~99 &     BC &  8.69   &  8.15    &  8.56  & C-rich \\
SMP~100 &    BC\textsuperscript{b}   &  8.55  &  7.61    &  8.36   & C-rich \\
SMP~101 &    BC\textsuperscript{b}   &$\dots$ & $\dots$  & $\dots$ & $\dots$ \\
SMP~102 &    BC &  8.65\textsuperscript{a}    & $\dots$  &  8.28   & F \\
\multicolumn{6}{l}{\textsuperscript{a}\footnotesize{Carbon abundance from \citet{letizia05}.}} \\
\multicolumn{6}{l}{\textsuperscript{b}\footnotesize{Uncertain morphology.}}
\end{tabular}
\end{table}

It is worth noting that LMC PN studies in such chemical detail, i.e.\ whose auroral 
lines have been observed for electron temperature detection, are at the bright end of the 
PN luminosity function (PNLF). Such bright PNe are believed to have progenitors  in the 
mid-mass range, given that the very high mass progenitors evolve too fast in the post-AGB 
to produce PNe that are observed at high luminosity. This translates into a very mild 
selection against the high AGB mass progenitors, something that is taken into account 
when interpreting the comparison between data and models.
It is also worth noting that extragalactic PNe are typically selected from [O~III] 
$\lambda$5007 off band-on band images. This may produce a selection effect toward higher 
oxygen abundance/ younger progenitors. In fact, at typical LMC metallicities, higher 
oxygen abundance seem to favour cooling through 
other emission lines than  [O~III] 
5007 $\AA$, contrary to what happens at higher metallicities (see Stanghellini et al.\ 2003).

\begin{figure*}
\begin{minipage}{0.49\textwidth}
\resizebox{1.\hsize}{!}{\includegraphics{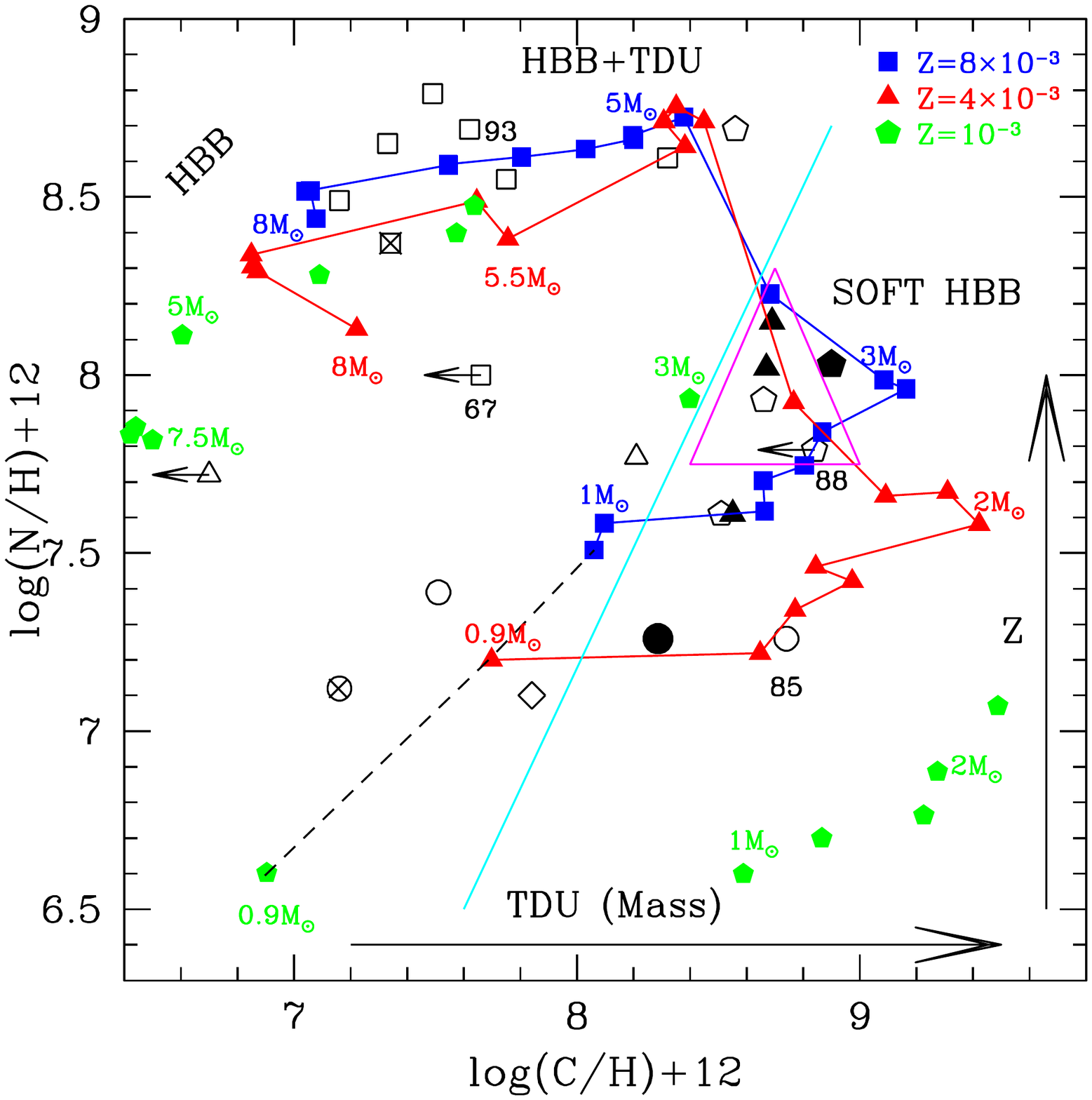}}
\end{minipage}
\begin{minipage}{0.49\textwidth}
\resizebox{1.\hsize}{!}{\includegraphics{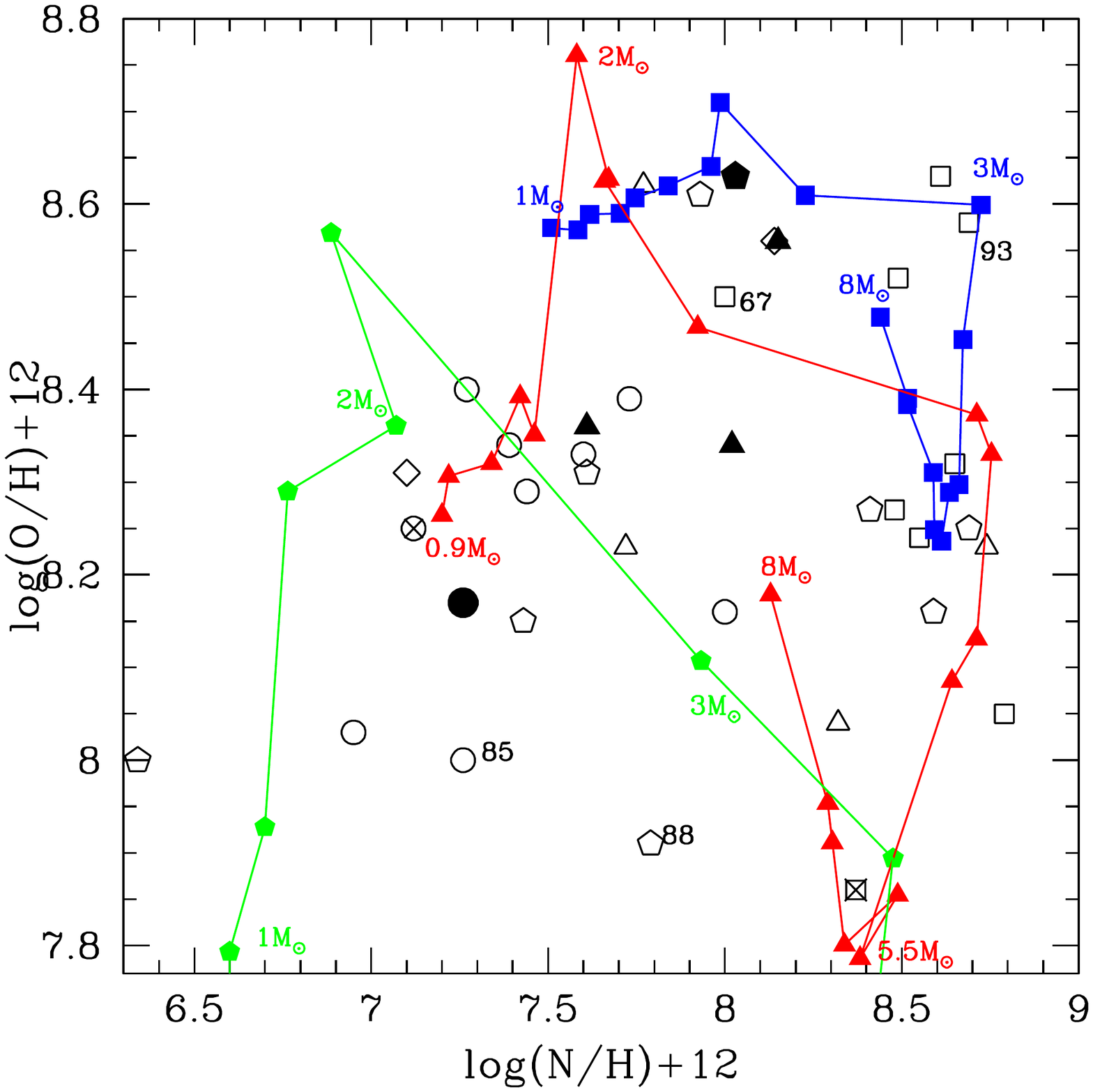}}
\end{minipage}
\vskip-50pt
\caption{Left: Stellar evolutionary models and PN data in the nitrogen  vs.\ carbon abundance plane. 
Carbon and nitrogen abundances of LMC PNe symbols are, according to 
their morphology, squares: bipolar; triangles: bipolar core; circles: round; pentagons: 
elliptical; diamonds: undetermined or uncertain morphology. The data symbols are filled 
where carbon dust, crossed  where  oxygen dust, has been detected.  
The loci of the AGB model final mass fractions are indicated with 
filled squares, triangles, and pentagons for $Z=8\times 10^{-3}$, $Z=4\times 10^{-3}$, 
and $Z=10^{-3}$ respectively. Models of different initial mass and the same metallicity are 
connected with solid lines; for clarity reasons, $Z=10^{-3}$ models are not joined with any
line. For the metallicities $Z=4,8\times 10^{-3}$ we show models
of mass in the range $1M_{\odot} \leq M \leq 8M_{\odot}$, with steps of $0.25M_{\odot}$
and $0.5M_{\odot}$ in the intervals, respectively, $1M_{\odot} \leq M \leq 2M_{\odot}$ 
and $2M_{\odot} \leq M \leq 8M_{\odot}$. 
In the $Z=10^{-3}$ case the overall mass interval
is $1M_{\odot} \leq M \leq 7.5M_{\odot}$, with steps of $0.25M_{\odot}$ in the low-mass
regime ($1M_{\odot} \leq M \leq 1.5M_{\odot}$) and $0.5M_{\odot}$ for masses above $2M_{\odot}$.
$0.9M_{\odot}$ models are also indicated for $Z=1,4\times 10^{-3}$. For some models the
initial mass of the precursor is indicated.
The dashed line indicates the locus of expected C and N 
abundances after the first dredge-up. The solid, diagonal (cyan) line separates the 
C-rich (right) from the O-rich (left) region. The triangular (magenta) region delimits 
the region populated by stars near the threshold mass to ignite HBB (see text for details). 
The horizontal large arrow indicates the in the low--mass regime the stars of higher
mass accumulate more carbon at the surface, owing to the higher number of TDU episodes
experienced; the vertical large arrow indicates that the surface nitrogen content
of low--mass AGBs increases with the metallicity of the star.
Right: Same plot, but in the oxygen vs.\ nitrogen abundance plane. Symbols are as in the 
left panel. In both plots, numbers indicate individual LMC PNe as in Table 1.
}
\label{fpne}
\end{figure*}

\subsection{Loci in the diagnostic diagrams}

The left panel of Fig.\ \ref{fpne} shows the observed PN abundances (by number, in the 
usual log($X$/H)+12 scale) in the CN plane. The final mass fractions of carbon and nitrogen, 
derived from the evolutionary models, have been plotted on the same scale.
In this plot the observed PNe are indicated with different symbols, depending on both their 
morphology and dust types. The model symbols have different colours and shapes depending 
on their metallicity. 

The correlation of PN properties with progenitor mass is analysed on the basis of the 
stellar evolution processes seen in the previous section. For a given metallicity, the 
chemistry of the lowest masses shown ($M \sim 1~M_{\odot}$) 
reflects essentially the effects of the first dredge-up, with a reduction of the initial
carbon and the increase in the surface nitrogen. Stars of higher mass (still below HBB 
activation) experience more TDU events, thus their final chemistry will be richer in 
carbon, whereas nitrogen keeps approximately constant.

For masses $M\sim 2.5~M_{\odot}$ the theoretical sequences bend leftwards. Models of mass
around the threshold limit to activate HBB (see discussion in section \ref{middle})
experience soft HBB, with a slight enhancement of the surface N, and partial destruction 
of the carbon previously accumulated. These models are included in the triangular region 
in the figure.

For $M > 3.5~M_{\odot}$ the final nitrogen is greatly increased by HBB. 
In agreement with the arguments discussed in section \ref{hbbmodel}, the most massive 
models 
occupy the upper-left side of the plane, as their surface chemical composition is not
contaminated by TDU, which favours very small carbon abundances. Models of lower
mass end their AGB evolution with a higher surface carbon, owing to the effects of a
few TDU episodes in the final stages, when HBB is extinguished.

In the CN plane the metallicity effect is more clearly seen in the low-mass regime, as 
these models evolve at approximately constant N, thus their position is sensitive to the 
initial chemical composition, particularly the original nitrogen when the star formed. 
In the high-mass domain the differences among the various metallicities are much smaller, 
because the C vs.\ N trend mainly reflects the equilibria of CN cycling.

As shown in Fig. \ref{fpne}, the models encompass observed groups of PNe. In particular, 
we note the dichotomy in the distribution of stars, divided into nitrogen-rich 
($\log (N/H)+12 > 8.3$) and nitrogen-poor ($\log (N/H)+12 < 8.2$) groups.

\subsection{Nitrogen-rich PNe}

By comparing the AGB final chemical composition to the observed LMC PN abundances we interpret that N-rich
PNe, most of which exhibit a bipolar morphology, 
are the progeny of stars with initial mass above $\sim 4~M_{\odot}$ that have experienced
HBB during their AGB evolution. The most 
carbon-poor PNe in the LMC ($\log (C/H)+12 < 7.5$) belong to this group 
and are the descendants of the most massive AGB stars, whose surface chemical composition 
reflects the effects of pure HBB. Conversely, PNe with a higher carbon (but that are still 
N-rich) correspond to the final phase of the evolution of smaller mass progenitors, whose 
surface chemistry was contaminated by both HBB and TDU.

Given the evolutionary time-scales of AGB stars within this range of mass, we conclude that 
this PN subsample has progenitors of mass above $\sim 4~M_{\odot}$, younger than $\sim$$200$ 
Myr. In particular, the nitrogen-rich and carbon-poor PNe have ages $\sim 40$--80 Myr and
descend from stars of mass in the range $6~M_{\odot} < M < 8~M_{\odot}$.

The analysis of the position of the N-rich sample in the CN plane does not allow us to
draw information on the PN progenitor's metallicity since the final chemistry
of stars experiencing HBB is fairly independent of the original chemical composition.
As shown in the left panel of Fig.\ \ref{fpne}, models of different metallicity overlap
with the observations.
The NO plane is more useful to this aim, because the final oxygen abundance is 
extremely sensitive to the initial metallicity. In the right panel of Fig.\ \ref{fpne}, 
we note that most of LMC PNe with enhanced nitrogen can be favourably compared with 
 $Z=8\times 10^{-3}$ models of mass in the range $4~M_{\odot} \leq M \leq 8~M_{\odot}$, 
with the exception of SMP21, an N-rich ($\log (N/H)+12 \sim 8.4$) PN surrounded by 
oxygen-rich dust. The low (gas) oxygen content of this PN ($\log (O/H)+12 \sim 7.85$) 
seems to indicate a low metallicity ($Z=4\times 10^{-3}$) precursor with initial mass 
$\sim$6--7$~M_{\odot}$. This interpretation is further confirmed by the Ar content of 
the PN (Leisy \& Dennefeld 2006), which is the lowest among the N-rich PNe observed.

\subsection{The progeny of low-mass AGB stars}
\label{crich}
The sample of LMC PNe shown in the CN plane includes PNe with $\log (N/H)+12 < 7.7$,
mostly with round or elliptical morphology. According to the comparison with our models, these
are the descendants of stars with mass below $\sim$$2.5~M_{\odot}$. The spread in the
observed nitrogen is an effect of a difference in the progenitor metallicity, higher-$Z$ stars having
higher nitrogen mass fraction. The observed spread in carbon is an indication of how many TDU episodes the progenitor star 
experienced during the AGB evolution; according to the discussion in section \ref{lowmass}, 
models of higher mass experience more TDU episodes so that they end their evolution with 
a higher content of carbon. The trend in carbon is, in this case, a trend in mass. Both the
effects of mass and metallicity are indicated, respectively, with a horizontal and a vertical,
large arrow in the left panel of Fig. \ref{fpne}.

PNe belonging to this low-carbon sub-sample are
the descendants of stars of mass $\sim 1~M_{\odot}$, showing contamination from
the first dredge-up only. The age of their progenitor stars is in the range of 5--10 Gyr. In the CN 
plane, they are close to the dashed line.  

Other carbon-rich PNe compare well with higher mass model ($\sim$1.5--$2.5~M_{\odot}$) yields 
and with progenitors formed between $\sim$500 Myr and $\sim$2 Gyr ago.

An additional group of LMC PNe show both carbon and nitrogen enhancement. They are enclosed 
within the triangular region in the left panel of Figure \ref{fpne}. We interpret these as 
the progeny of stars with mass close to the threshold value to activate HBB 
($M \sim 3~M_{\odot}$) discussed in section \ref{middle}. The enhancement in nitrogen is 
due to weak HBB experienced in the AGB phases prior to the C-star stage, after which 
the stars have undergone a series of TDU episodes that increased the surface carbon.

\subsection{A few outliers}
The comparison among the observed abundances and the surface chemical composition in the
final evolutionary phases allows an
interpretation of most LMC PNe whose CNO abundances are known in terms of the mass (i.e.\ age) and
metallicity of the progenitors. 

Nonetheless, there are a few outliers. Four PNe apparently fall outside the range of CNO abundances
covered by the models and thus deserve dedicated analysis. These PNe, indicated in the
CN plane with the corresponding SMP numbers, are SMP~67,
SMP~85, SMP~88, and SMP~93. 

\begin{itemize}

\item{SMP~67: The position of this PN in the CN plane is apparently not encompassed
by any model. However, the carbon abundance observed for SMP~67 is an upper limit, based on 
non-detection (see Leisy \& Dennefeld 2006 for details), opening 
the possibility that this is the descendant of a massive AGB stars that experienced HBB 
with little (or no) contamination from TDU. The bipolar morphology also points to this 
interpretation.
The oxygen content, as shown in the right panel of Fig.\ \ref{fpne}, is compatible with the 
$Z=8\times 10^{-3}$ metallicity.
}

\item{SMP~85. From the position in the CN plane (see left panel of Fig.\ \ref{fpne}), 
this round-shaped PN could be the descendent of a low--mass star of $Z=4\times 10^{-3}$ 
that experienced some carbon enrichment, owing to the effects of a few TDUs. 
The observed low oxygen abundance is hard to compare with any of the models. 
}

\item{SMP~88: The main problem in the interpretation of the chemistry of this
PN is its low oxygen content, which rules out the possibility
that it is the descendant of a low-mass star of metallicity $Z=4$--$8 \times 10^{-3}$,
as deduced from its position in the CN plane. Taking into account that carbon was 
not detected for this object (an upper limit in given based on the spectrum), and 
considering a typical error on the oxygen content of $\sim 0.1$ dex, we suggest that this 
PN originates from a low-metallicity star of mass $\sim 3~M_{\odot}$, falling in the sample 
of AGB stars undergoing modest HBB and then becoming carbon star. This hypothesis is 
confirmed by the very small Ar content ($\log (Ar/H)+12 = 5.61$).
}

\item{SMP~93: In the CN plane, this bipolar PN is compatible with being the product of 
evolution from an AGB star that underwent 
HBB and (possibly) of some TDU processing. The problem in the interpretation of this 
object is its position in the NO plane, where the high oxygen abundance is compatible 
with a low-mass progenitor, a progenitor that experienced soft HBB and several TDU episodes. 
If the oxygen abundance observed is overestimated by $\sim 0.2$ dex, the chemistry of the star
would be then compatible with its being the descendant of a massive AGB of metallicity 
$Z=4\times 10^{-3}$.
}

\end{itemize}

\section{PNe in the LMC as probes of stellar evolution theory}

The dichotomy in the distribution of PNe in the CN plane, particularly the division
among N-rich ($\log (N/H)+12 > 8.3$) and N-poor objects, confirms the existence of a 
threshold in mass, above which the stars experience HBB and become enriched in nitrogen. In 
the present analysis we find that HBB is active in all stars of initial mass above 
$\sim$$3~M_{\odot}$. This limit is partly dependent on convection modelling and on the 
assumptions concerning the overshoot from the convective core during the core H-burning 
phase; other research groups find a slightly higher threshold mass ($\sim$$4~M_{\odot}$) at 
the same metallicities discussed here \citep{karakas10}. 

The existence of a group of nitrogen-rich and carbon-poor PNe indicates that the surface
chemistry of their precursors have been contaminated mainly by HBB. These stars exhibit 
a pure HBB chemistry and, according to our interpretation, they descend from very massive 
($M > 6~M_{\odot}$) AGB stars, undergoing a small number of weak thermal pulses. While 
some effects of TDU during the AGB evolution cannot be disregarded, the observed, small 
abundances of carbon rule out the possibility that any TDU episode occurred in the
very final AGB phases, when HBB was no longer active: indeed, this would determine an
increase in the carbon mass fraction. 

This finding offers an important opportunity to discriminate among the models of massive 
AGB stars present in the literature. The dissimilarities among the results presented by
various research groups is an indication of the uncertainties affecting the modelling
of these stars, mainly owing to the extreme sensitivity of the results obtained concerning the 
description of convection and mass-loss \citep{ventura11, ventura13, doherty14b}. 
\citet{doherty14a} outlined the relevant effect of the treatment of the 
convective boundaries on the possible occurrence of TDU in massive AGBs and concluded that 
the differences among the carbon yields between their models and those presented 
by \citet{siess10} \citep[see Fig. 15 in][]{doherty14a} were due to the different methods 
used to determine the extent of the mixed region at the base of the convective mantle. The 
presence of PNe with very small carbon contents is an indication that the chemistry of 
massive AGBs is not (or only scarcely) contaminated by TDU, in agreement with 
\citet{siess10} and \citet{ventura11}. 

All nitrogen-rich PNe studied here have C/O $<$1. Understanding whether this result is 
limited to the sample of PNe studied here, or whether it can be generalized to all the PNe in 
the LMC, would be crucial to assessing the variation of the surface chemical composition 
of AGBs with mass 4--6$~M_{\odot}$, that experience soft HBB, with temperatures below $\sim$80 MK. 
\citet{frost98} suggested that these stars would eventually reach the C-star stage
as a consequence of a series of TDU episodes after HBB was shut down by the loss of the 
external mantle. The results of \citet{karakas10}, based on full evolutionary computations,
confirmed this possibility. 
Conversely, results presented by \citet{ventura13} indicate that the C-star stage is never reached 
by models experiencing HBB, which leaves no room for the possibility of observing PNe enriched 
in nitrogen and with C/O above unity. The observation of some PNe in the upper region of 
the CN plane, on the right of the line separating oxygen-rich objects from carbon-rich PNe
(see left panel of Fig.\ \ref{fpne}), would confirm the mechanism suggested by \citet{frost98} 
for forming bright carbon stars.  

Still in the context of models experiencing HBB, we note that while the chemical
composition in the final evolutionary phases of massive AGB stars 
of metallicity $Z=4, 8 \times 10^{-3}$ are in good agreement with the observed PN abundances
of C, N and O, the same does not hold for the low-$Z$ component. The reason
is that these stars, within the present modelling of convection based on the FST
treatment, are expected to experience strong HBB \citep{vd05}, with a considerable 
depletion of the surface oxygen (see right panel of Fig.\ \ref{fz1m3}); this is at odds 
with other models in the literature \citep{karakas10}, where the oxygen destruction 
(if any) is scarce. The right panel of Fig.\ \ref{fpne} 
shows that such extremely low abundances of oxygen are indeed not found in the present 
sample of PNe. A possible explanation for this is an observational bias towards the PNe 
with high oxygen, which is possible by having selected  PNe that are $\lambda$5007-bright and 
thus, at low metallicity, typically oxygen-rich \citep{letizia03}. Alternatively, 
the HBB produced by low-metallicity, massive AGB models, where convection is treated within 
the FST framework, is overestimated.

The detection of carbon-enriched PNe that are partly nitrogen-enriched confirms the mechanism
of quenching of HBB, favoured by the achievement of the C-star stage. The presence of
these PNe further supports the necessity of using the low-temperature, carbon-rich opacities 
when modellng these evolutionary phases, as pointed out by \citet{marigo02, marigo07}.

Following the discussion in section \ref{crich}, the PNe on the lower-right side of the
CN plane are interpreted as the progeny of stars with initial mass below the threshold 
needed to activate HBB. The observations indicate an upper limit to the carbon abundance 
($\log (C/H)+12 < 9$), which corresponds to a surface mass fraction $X({\rm C}) \sim 10^{-2}$. 
This is in agreement with our predictions, with the exception of models of mass 
$\sim$2--$2.5~M_{\odot}$, that reach slightly higher surface carbon in the late evolutionary 
AGB phases. 

In low-mass AGBs, the final surface carbon is the outcome of a series of TDU events that 
gradually enrich  the external mantle in carbon. The result depends on a delicate interplay
between the rate at which mass is lost and the extent of TDU: the former determines the 
number of thermal pulses (and hence of TDU episodes) experienced by the stars during the
AGB phase, whereas a more efficient TDU favours a faster increase in the surface carbon.
The upper limit given above, if confirmed, indicates that once the 
C-star stage is reached, the external mantle is lost rapidly, thus limiting the
number of thermal pulses experienced; radiation pressure acting on solid carbon dust 
particles could be a possible explanation for the increase in the mass-loss rate during 
these phases. This is in agreement with the AGB models used in the present investigations
and with other models in the literature predicting only a modest increase in the surface
carbon, with a final C/O ratio below $\sim$4 \citep{weiss09}.

On the other hand, AGB models by other groups predict much higher carbon enrichments,
with C/O ratios up to $C/O \sim 10$ \citep{karakas07, cristallo11}. The results obtained
here would suggest that the extent of the TDU experienced by carbon stars is lower than 
predicted by the latter models and/or that mass is lost much faster from the external
envelope. To discriminate among the various models further observations of C-rich
PNe in the LMC are required.

\section{Conclusions}
We analysed the LMC PNe whose direct CNO abundances are available from space- and 
ground-based observations, and compared them with the chemical composition of stars
of mass $1~M_{\odot} < M < 8~M_{\odot}$ based on AGB evolution,
accounting also for dust formation in the circumstellar envelope. 

The observed abundances of carbon, nitrogen, and oxygen of the PNe sample encompass
the final surface chemical composition of the AGB models that we used for the comparison, 
which allows us to characterize the PNe sample in terms of age, chemical composition, 
and mass of the progenitors.

The chemical composition of the PNe exhibits a dichotomy in the distribution of the
nitrogen abundances, which we interpret as due to the activation of HBB
in the stars of initial mass above $\sim$$3~M_{\odot}$. The objects enriched in nitrogen, 
with $\log (N/H)+12 > 8$, are the progeny of stars with mass higher than the aforementioned 
threshold, formed 50--200 Myr ago. The spread in the carbon content of these nitrogen-rich 
PNe is explained by the different number of TDU episodes experienced, particularly 
in the very late evolutionary phases, when HBB is shut down by the envelope consumption. 

The N-rich PNe with the lowest carbon abundance are interpreted as the progeny of massive AGB stars
with $M \geq 6~M_{\odot}$, whose surface chemistry reflects mainly the effects of HBB, 
with a modest contamination from TDU. The very low carbon abundances of these stars 
confirms that efficient HBB occurs at the base of the envelope of massive AGB stars, in 
agreement with the predictions of the FST modelling of convection.
The lack of carbon stars among the N-rich group suggests that formation of bright 
C-stars via repeated TDU episodes following HBB quenching does not occur: this is in 
agreement with our modelling, though additional measurements of carbon abundances of LMC PNe 
are required to clarify this issue further.

PNe with $\log (N/H)+12 < 7.5$ are interpreted as the final stages of stars that did 
not experience any HBB. These objects, with initial mass below $\sim$$3~M_{\odot}$,
formed between 500 Myr and 10 Gyr ago. Whereas the spread in the carbon measured originates
from the different number of TDU episodes experienced, the variation in the nitrogen
content reflects the difference in the metallicity.

The highest carbon abundances found in the present sample of PNe suggests an upper 
limit to the amount of carbon accumulated at the surface of these stars, with 
$\log (C/H)+12 < 9$. This finding, if confirmed, indicates that once the C-star phase
is reached, the stars lose their external mantle very rapidly, thus limiting the
number of further TDU episodes experienced. Other observations are needed to confirm
this conclusion.

\section*{Acknowledgments}
We are particularly indebted to the anonymous referee for the careful reading of the
manuscript, that helped improving the quality of the paper.
P.V. was supported by PRIN MIUR 2011 "The Chemical and Dynamical Evolution of the Milky Way 
and Local Group Galaxies" (PI: F. Matteucci), prot. 2010LY5N2T. D.A.G.H. 
acknowledges support provided by the Spanish Ministry of Economy and Competitiveness under 
grant AYA-2011-27754. M.D.C acknowledges support from the Observatory of Rome.

\end{document}